\documentclass[10pt]{article} 
\hoffset = - 0.5 truecm 
\def\beq{\begin{equation}}   
\def\eeq{\end{equation}}
\def\bea{\begin{eqnarray}}  
\def\eea{\end{eqnarray}}

\def\beeq{\begin{eqnarray}} 
\def\eeeq{\end{eqnarray}}
\def\lsim{\raise0.3ex\hbox{$<$\kern-0.75em\raise-1.1ex\hbox{$\sim$}}}
\def\gsim{\raise0.3ex\hbox{$>$\kern-0.75em\raise-1.1ex\hbox{$\sim$}}}

\usepackage{graphicx}

\usepackage{epsfig}

\newcommand\mysection{\setcounter{equation}{0}\section}

\renewcommand{\theequation}{\thesection.\arabic{equation}}

\newcounter{hran} \renewcommand{\thehran}{\thesection.\arabic{hran}}

\def\bmini{\setcounter{hran}{\value{equation}}

	 \refstepcounter{hran}\setcounter{equation}{0}

	 \renewcommand{\theequation}{\thehran\alph{equation}}\begin{eqnarray}}

\def\bminiG#1{\setcounter{hran}{\value{equation}}

\refstepcounter{hran}\setcounter{equation}{-1}

\renewcommand{\theequation}{\thehran\alph{equation}}

\refstepcounter{equation}\label{#1}\begin{eqnarray}}

\def\emini{\end{eqnarray}\relax\setcounter{equation}{\value{hran}}\renewcommand{\theequation}{\thesection.\arabic{equation}}}

\begin{document}

\begin{titlepage}

 \begin{flushright}

	   \today\\

	   arXiv:xxxx.yyyy [hep-ph]\\

	   LAPTH-1259/08 \\

	   IMSc-2008/06/11\\

	   LPT-Orsay-08-59\\

\end{flushright}

\vspace{1.cm}

\begin{center}

\vbox to 1 truecm {}

{\large \bf Jet-jet and hadron-jet  correlations in hadro- and
electro-production}

\par \vskip 3 truemm

{\large \bf   }

\vskip 1 truecm

\centerline{\bf {P. Aurenche$^{1}$, Rahul Basu$^{2}$, M. Fontannaz$^{3}$}}

\vskip .5cm

{\small\sl

\centerline{ $^1$LAPTH{\footnote{Laboratoire d'Annecy-le-Vieux de Physique
 Th\'eorique, UMR 5108 du CNRS.}}, Universit\'e de Savoie, CNRS,}


\centerline{B.P.110, F-74941 Annecy-le-Vieux Cedex, France}

\vskip .5cm

\centerline{ $^2$The Institute of Mathematical Sciences,}

\centerline{Chennai 600 113, India}

\vskip .5cm

\centerline{ $^3$Laboratoire de Physique Th\'eorique, UMR 8627 CNRS,}
\centerline{Universit\'e Paris XI, B\^atiment 210, 91405, Orsay,
Cedex, France}
}


\vskip 2 truecm

\begin{abstract} We discuss, in the framework of perturbative QCD
at next to leading order, two related observables which are usually
considered to provide tests of the BFKL dynamics : jet-jet correlations
at Tevatron energies and forward particle-jet correlations at HERA.
In the first case we study the rapidity gap dependence of the
azimuthal correlations and find slightly too strong correlations
at large gap. In the second case we discuss the cross section as
well as the azimuthal correlations over a rapidity gap range of 5
units. We find that the requirement of a forward particle imposes
strong kinematical constraints which distort the distributions,
notably at small rapidity gaps. We also show that the decorrelation is stronger
in electroproduction than in hadron-hadron collisions. Unfortunately no data are yet
available for comparison.  
\end{abstract}

\end{center}

\end{titlepage}

\pagestyle{plain}

\baselineskip=22 pt

\mysection{Introduction} 
\hspace*{\parindent} 
The first test of BFKL~\cite{Lipatov:1976zz} dynamics in hadronic collisions was
proposed by Mueller and Navelet~\cite{Mueller:1986ey} who showed that the cross
section for the inclusive production of 2 jets at the partonic level
would increase exponentially with the rapidity separation, $\Delta \eta$,
of the jets. At the hadronic level the cross section involves a
convolution with the initial parton distributions, as a result of which
the hadronic cross section decreases with the rapidity gap due to the
very fast fall-off of the parton densities. In order to disentangle the
interesting dynamical effects from the parton distributions effect 
the study of the normalized azimuthal angle
distribution~\cite{DelDuca:1993mn,Stirling:1994zs} was proposed which should be less sensitive to the latter.
More precisely, the relevant quantity to consider is the average $<\cos
(\pi - \phi ) >$, where
$\phi$ is the azimuthal angle between the 2 jets. From BFKL dynamics one
would expect a very fast decorrelation, due to the emission of a growing
number of jets in the rapidity interval between the observed jets. This
is to be contrasted to the case of DGLAP~\cite{Gribov:1972ri} dynamics as
implemented in lowest order (LO) and next-to-lowest order (NLO)
calculations: at LO the 2 jets are perfectly correlated ($\phi = \pi$)
while at NLO the decorrelation arises from the emission of just one
extra  jet.  Monte-Carlo codes based on DGLAP dynamics offer however the
possibility of a stronger decorrelation due to parton showers.

Coming back to BFKL dynamics it was noted~\cite{4r,Schmidt:1996fg,Orr:1997im} that
the predictions were very sensitive to energy momentum constraints
neglected in the original analytic calculations.  This is, in
particular, true for azimuthal correlations where an exact energy
momentum conservation considerably softens the rapidity dependence,
compared to the analytical (asymptotic) results. Indeed, less phase
space is available for the emission of intermediate gluons in the
former case. Because of these and other effects~\cite{Orr:1997im,Vera:2007kn} the
BFKL predictions show less decorrelation than that found in
the initial calculations.

Data on $<\cos (\pi - \phi ) >$ at Tevatron energies have been obtained
by the D\O\ collaboration~\cite{Abachi:1996et} in the range $0 < \Delta \eta < 5$
and have been compared to a variety of calculations. The leading order
BFKL asymptotic predictions are completely ruled out by the data while
the leading order ones, including energy momentum
conservation~\cite{4r,Orr:1997im}, or the next-to-leading order
ones~\cite{Vera:2007kn} are closer to the data with still a too strong
rapidity decrease. On the other hand, the NLO predictions of
JETRAD~\cite{Giele:1994gf} do not show enough decorrelation as the rapidity
increases. Only the HERWIG~\cite{Marchesini:1987cf} results are in very good agreement
with the data.

BFKL dynamics can also be probed by considering the cross section
for forward production of a large $p_{\bot}$ jet or particle in
deep-inelastic scattering (DIS) events at small Bjorken-$x$ $x_{Bj}$~\cite{Mueller:1990er},
in the domain $p_{\bot}^2 \sim Q^2$. Such data have been collected
by the H1~\cite{Adloff:1998fa,Aktas:2004rb} and ZEUS~\cite{Breitweg:1998ed}
collaborations. For this observable, models based on BFKL dynamics
(supplemented by some phenomenological inputs)~\cite{Kwiecinski:1999hv}, were
claimed to be in reasonable agreement with data although they did
not account for the correct $Q^2$ evolution. More recently, several
NLO calculations of forward $\pi^0$ production have been carried
out~\cite{Aurenche:2003by,Fontannaz:2004ev,Aurenche:2005cb,Daleo:2003xg,Daleo:2004pn,Kniehl:2004hf} and found in
agreement with the latest H1 data~\cite{Aktas:2004rb}: in particular,
the dependence of the cross section as a function of the variables
$x_{Bj},\ Q^2, \ x_{\pi}, p_{\bot}^*$ (where $x_{\pi}$ and $p_{\bot}^*$
are, respectively, the fraction of longitudinal momentum and the
transverse momentum of the pion in the $\gamma^*$-proton center of mass
frame) are very well reproduced~\cite{Aurenche:2005cb}.

To further study the validity of the perturbative approach in
deep-inelastic scattering, we propose to consider azimuthal correlations
involving a forward large $p_{\bot}$ hadron, to probe the small
$x_{Bj}$ where BFKL dynamics is often thought to be more relevant.
Because of the different subprocess structure, it is interesting
to compare the azimuthal correlations in hadroproduction and in
electroproduction.  However, in the latter case, the perturbative
dynamics is more complex because  the cross section is built from
two contributions, the so-called direct and resolved contributions.

In the following section, we first go back to correlations in $p\ \bar p$
collisions at the Tevatron, from a perturbative point of view, and find
that the rapidity dependence, quite sensitive to scale changes, is within
the experimental errors, although being not steep enough. In
Sect.~\ref{azim-correl-DIS} we turn to DIS and study forward-pion jet
correlations. We consider two classes of observables: the cross section
and the average $ < \cos(\pi - \phi) > $ as a function of the rapidity hadron-jet
separation for different $Q^2$ values. The importance of kinematical
constraints on the shape of the rapidity distribution is stressed.
Conclusions are given in a final section.

\mysection{Azimuthal correlations in hadron-hadron collisions}
\hspace*{\parindent} 
The D\O\  collaboration measured~\cite{Abachi:1996et} the
rapidity dependence of the azimuthal correlations between
large-$p_{\bot}$ jets at $\sqrt{s} =$~1800~GeV with the following
criteria:  among the jets with $E_{\bot} > 20$~GeV, the two jets
having the largest and the smallest pseudorapidity were selected.
One of the two jets was also required to be above 50 GeV. The
selected jets are required to be in the pseudorapidity range $|\eta_{1,2}|
\leq 3$ and the azimuthal correlations are measured as a function
of $\Delta \eta = \eta_1 - \eta_2$. More precisely, D\O\  measures
$<\cos (\pi - \phi)>$ as a function of $\Delta \eta$ where $\phi$
is the azimuthal angle between the two jets.
\vspace{1cm}

\begin{figure}[htb] 
\centering
\includegraphics[height=3in]{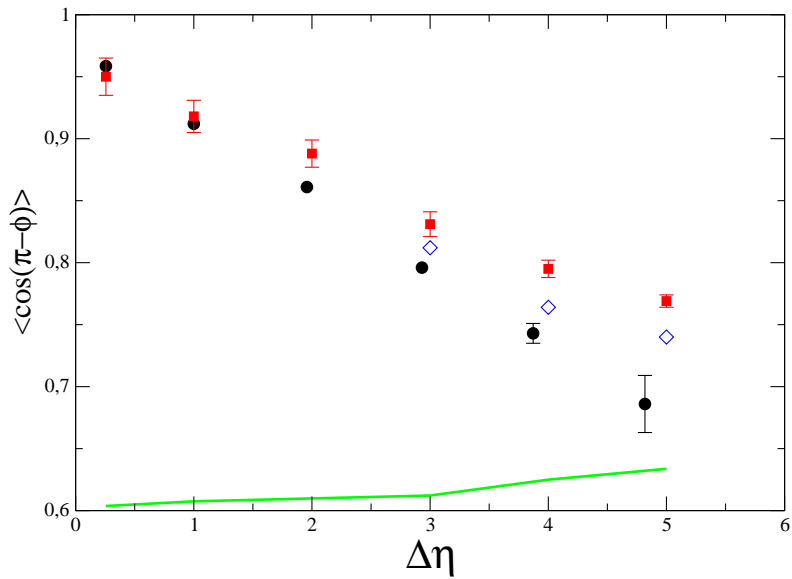} \caption{Theoretical
predictions, at NLO in perturbative QCD, of the two-jet correlations
(squares: scale $P_{\bot}/2$; diamonds: scale $P_{\bot}/4$)
compared with D\O\ data~\cite{Abachi:1996et} (solid dots). The solid line
indicates the systematic errors.} 
\label{fig:1} 
\end{figure}

To study this quantity we use the code DIJET which belongs to the PHOX
Family \cite{12r} and calculate single jet and dijet cross sections at NLO
in hadron-hadron collisions. The interested reader may find technical
details on the PHOX Family codes on the site \cite{12r}. We have
checked that the DIJET predictions are in good agreement with the single
jet cross section of ref.~\cite{Abbott:1998ya} and with the spectrum $dN/d\Delta
\eta$ measured by D\O\ \cite{Abachi:1996et} under the conditions described above. The
parton distribution functions are CTEQ6M \cite{Pumplin:2002vw} and we use
$P_{\bot}/2$ as the standard factorization and renormalization scales,
where $P_{\bot}$ is half the sum of the transverse momenta of the largest
$p_{\bot}$ partons,  $P_{\bot} = {p_{\bot 1} + p_{\bot 2} \over 2}$. For
the jets we use $R_c = .7$ and $R_{sep} =1.3 \ R_c$ as in ref.
\cite{Abbott:1998ya}.

We now focus on the azimuthal correlations. We calculate the values
of $<\cos (\pi - \phi )>$ quoted by the D\O\ collaboration in Fig.~4
of ref. \cite{Abachi:1996et}. We expect the corresponding theoretical predictions
to be sensitive to the choice of scales. Indeed the aplanarity of
the jets is due to the $2\to 3$ subprocesses and when $\phi$ is
different from $\pi$ we perform a LO calculation. Only the region
$\phi \sim \pi$ corresponds to a full NLO calculation. To test this
sensitivity we start with  the standard scale ${P_{\bot}/2}$. The
result is displayed in Fig.~\ref{fig:1} (squares) together with the
D\O\ data (solid dots). The theoretical error due to the limited
statistics in the Monte Carlo integration is of the order of $\pm
\ 10\  \%$ or less.

Another scale which may appear more natural is (minus) the square
of the momentum transfer in the $t$-channel of a $2 \to 2$ subprocess,
$Q^2 = p_{\bot}^2 (1 + e^{-(\eta_1 - \eta_2)})$, $p_{\bot}$ being
a jet transverse momentum. In order to have a scale close to
$p_{\bot}/2$ when $\eta_1 \sim \eta_2$ we use ${c}\ P_{\bot} \left
( {1 + e^{-(\eta_1 - \eta_2)}\over 2} \right )^{1 \over 2}$ with
$c = {1 \over 2}$ for the factorization and the renormalisation
scales. We obtain results almost identical to those obtained with
the scale ${P_{\bot} / 2}$ (about 1 \% lower). Finally with the
scale ${P_{\bot} / 4}$ we get the results shown by the diamonds in
Fig.~\ref{fig:1}.

The agreement between data and theory is not perfect, the theoretical
points staying at the upper limit of the experimental error
(statistical + systematic). The smooth decrease of $<\cos (\pi -
\phi )>$ with increasing $\Delta \eta$ is well reproduced at small
and medium rapidity gaps, but at large $\Delta \eta$ the jets in
the NLO calculation appear to be too much correlated. Note that
with the scale $\sum_{i=1}^3 p_{\bot i}/2$ we recover the JETRAD
results quoted in~\cite{Abachi:1996et}.







Other approaches based on the generator Herwig \cite{Marchesini:1987cf} and on a
BFKL calculation \cite{4r} are also displayed by the D\O\ collaboration.
The Herwig results are in good agreement with data whereas the BFKL
prediction undershoots them by a large amount. A recent BFKL analytic
calculation at next to leading order~\cite{Vera:2007kn} gives a somewhat
better description than the leading order one of Del Duca and Schmidt
quoted in \cite{Abachi:1996et} which takes into account some non asymptotic
effects.

The conclusion of this study is that HERWIG, and to a lesser extent
NLO calculations, can give a satisfactory description of the data,
although the underlying dynamics is different. In the NLO calculation
the jet aplanarity comes from the $2 \to 3$ subprocesses described
by exact QCD matrix elements. These matrix elements are not present
in HERWIG where the aplanarity comes from the initial and final
parton showers. This mechanism leads to a correct value of $<\cos
(\pi - \phi )>$. (However the cross section for three large-$p_{\bot}$
jets is expected to fall an order of magnitude below data, because
of the lack of exact $2 \to 3$ matrix elements in this approach).
Based on this analysis, we speculate that a NNLO calculation
would give a satisfactory description of the data.

\section{Hadron-jet azimuthal correlations in DIS} 
\label{azim-correl-DIS}
\hspace*{\parindent} In ref.~\cite{Aurenche:2005cb} we carried out a detailed
study of single inclusive forward $\pi^0$ production in deep inelastic
scattering and compared the NLO perturbative results with the recent
H1 data (see ref.~\cite{Adloff:1998fa,Aktas:2004rb} for experimental details
and cuts). The data probe a small $x_{Bj}$ range, $1.1\ 10^{-4} <
x_{Bj} < 1.1\ 10^{-3}$ where DGLAP dynamics may break down. The
calculation includes the ''direct" component (where the virtual
photon couples to the hard processes) and the ''resolved" one where
the photon acts as a composite object, both contributions being
calculated at the next to leading order. We use the FORTRAN code
DISPHOX developed from the real photoproduction code JETPHOX \cite{12r} 
and described in refs.~\cite{Aurenche:2003by,Fontannaz:2004ev,Aurenche:2005cb} with the following inputs:
the proton structure functions are taken from the CTEQ6M
tables~\cite{Pumplin:2002vw}, the parton fragmentation functions into pions
are those of KKP~\cite{Kniehl:2000fe} while, for the resolved case, the photon
structure functions are found in~\cite{Fontannaz:2004ev}. We work in the
$\overline{\rm MS}$ renormalization and factorization schemes and
all the scales are equal to $(Q^2 + p_{\bot}^2)$. We take $n_f =
4$ flavors and  $\Lambda_{\overline{MS}} = 326$~MeV.



\begin{figure} 
\vspace{9pt} 
\begin{center}
\includegraphics[width=15cm]{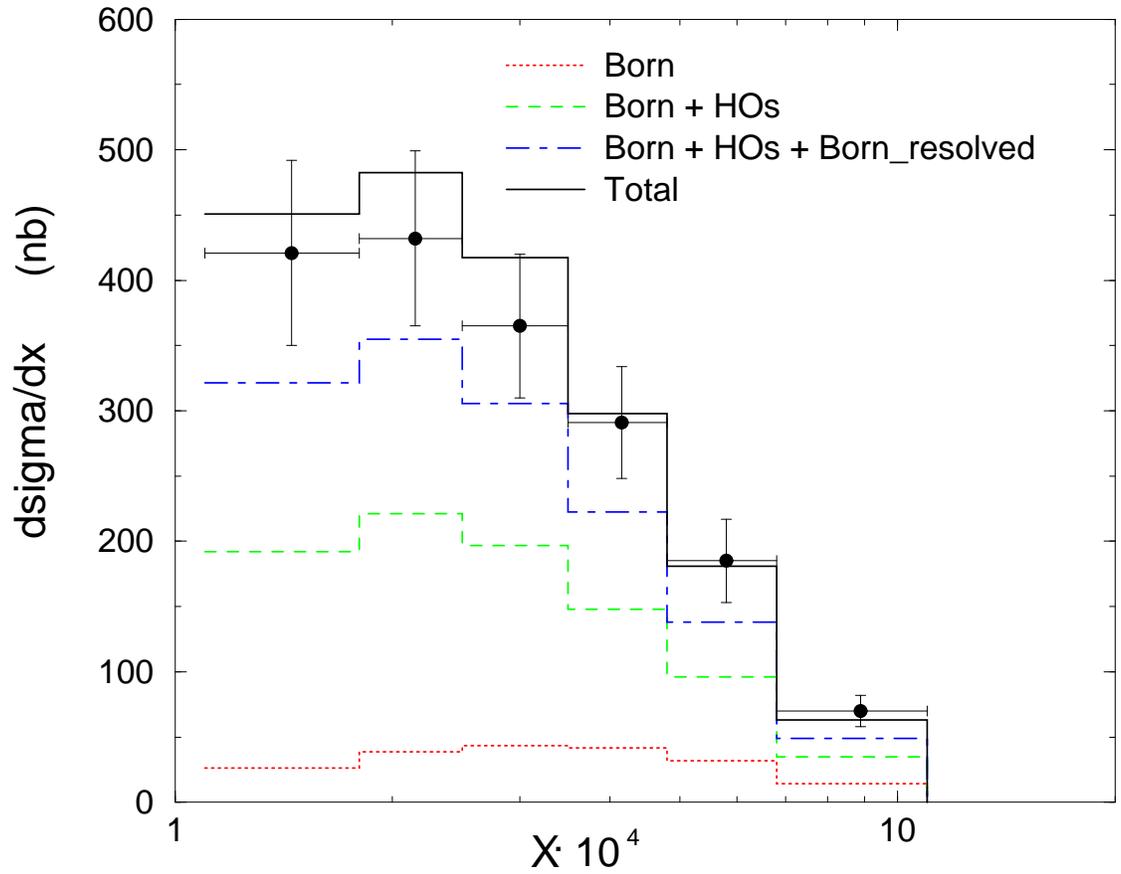} 
\end{center} 
\caption
{The cross section $d\sigma/dx_{Bj}$ corresponding to the range
4.5~GeV$^2 \leq Q^2 \leq 15$~GeV$^2$ compared to H1 data \cite{Adloff:1998fa}.
The various theoretical components of the cross section are shown.
The symbol HO$_s$ denotes the HO correction from which the lowest
order resolved contribution has been subtracted.} 
\label{fig:2}
\end{figure}

\begin{figure} 
\vspace{9pt} 
\begin{center}
\includegraphics[width=3.5in,height=2.5in]{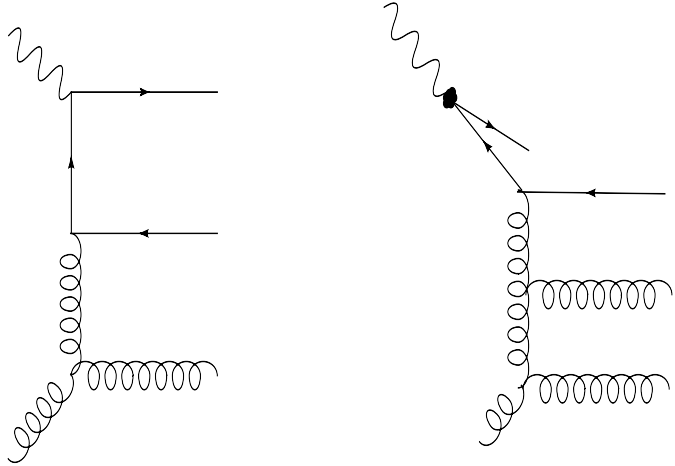} 
\end{center}
\caption {Examples of higher order corrections to (a) the direct
contribution, (b) the resolved contribution. The forward hadron is
emitted from the bottom parton line.} \label{fig:3} 
\end{figure}

In Fig.~\ref{fig:2} we display the single inclusive forward pion
distribution as a function of $x_{Bj}$ and its comparison with the
various theoretical components of the cross section. A remarkable
fact is the smallness of the Born direct term and the very large
size of the higher order correction to this Born term (see \cite{Aurenche:2005cb}
for a precise definition of various components). As already noted,
at lowest order, the hard processes are mediated by quark exchange
while, at higher orders, there appear diagrams with a gluon exchange
in the $t$ channel, {\it e. g.} $\gamma q\rightarrow q q \bar q$:
because of the gluon pole these terms are enhanced in the case of
forward production of hadrons and they  considerably increase the
production rate. The importance of the resolved component can also
be explained along the same lines because, already at lowest order,
it contains processes with a gluon exchange. In that case, the
hard processes, at LO and NLO, are the same as in $p \bar p$
collisions, the only difference with the latter case being the
convolution with the structure functions: those of the photon are
less steep than those of the proton and the relative weight of gluon
and quark initiated processes are different. In Fig.~\ref{fig:3},
we show two diagrams characteristic of the higher order corrections
included in our calculation: it is clear that these diagrams can
be seen as the lowest order terms of the BFKL ladder stretched
between the photon (backward direction) and the proton (forward
direction). If the available phase space (rapidity) between the
forward and backward partons is large enough then multiple gluon
emission is expected and the BFKL dynamics should dominate. In the
opposite case the emission of one (or two) gluons is probably more
appropriate. The very good agreement between the inclusive H1
data~\cite{Aktas:2004rb} and our perturbative calculation~\cite{Aurenche:2005cb}
seems to support the latter hypothesis which can be further tested
by considering forward hadron-jet azimuthal correlations. 



The H1 collaboration has studied azimuthal correlations between
the highest two transverse momentum jets~\cite{Aktas:2004px} in
DIS events at small $x_{Bj}$. They find a reasonable agreement with
perturbative calculations except for small separations in azimuth
where models based on BFKL dynamics fare better. Unlike the analysis
in $p\ \bar p$ collisions they do not study the correlations between
largest rapidity gap jets nor do they measure $<\cos (\pi - \phi)>$ as a
function of the rapidity gap.

In the following we study particle-jet correlations and we use the
same DISPHOX code which was successfully used to describe the single
inclusive forward pion data and we concentrate on the rapidity
dependence of cross sections and $<\cos (\pi - \phi)>$. 
Electron-jet correlations have already been considered 
in~\cite{{Bartels:1996wx},{Vera:2007dr}} in
the framework of the BFKL approach. In contrast  
to~\cite{{Bartels:1996wx},{Vera:2007dr}}, we propose to
measure the forward hadron-jet correlations in which the trivial effect
of the electromagnetic vertex is not present. These correlations are also
closer to those measured in hadron-hadron collisions.

The momentum of a large-$p_\perp$ final state is
better fixed in experiments measuring hadrons than in those measuring
jets. In the latter case, the link between jet and parton momenta
are obtained via Monte-Carlo calculations and it might not be well
controlled for small transverse momentum jets, with $p_{\bot} <
4$~GeV/c say. Besides, using a particle rather than a jet also
allows us to extend the useful rapidity range of the observable.

All our results will be presented in the hadronic center of mass
frame (CM$^*$) where the virtual photon and initial parton in the
proton are aligned along the $z$ axis. The reason is the following.
In the laboratory frame the photon has a transverse momentum
($\vec{q}_{\bot}^{\ 2} = (1 - y)Q^2$){\footnote{$y$ is the inelasticity of
the electron $y =  q\cdot P/ P_e \cdot P$, with $P_e$ and $P$ the initial      
electron and proton momenta respectively.}}  and therefore it can emit a
collinear parton which can be part of a large-$p_{\bot}$ jet. In
the resolved case this parton is also collinear to the spectator
partons contained in the photon and these latter must also be part
of a large-$p_{\bot}$ jet. As a result, large-$p_{\bot}$ jets defined
in the laboratory frame can be made of spectator partons, and we
would need to modify the jet algorithm used in DISPHOX which only
considers jets made of active partons (produced in the subprocess).
Besides, the cross section would be very sensitive to the position of the 
cone (defining the jet) with respect to the photon momentum. For instance, a cone
tangent to the photon momentum will lead to a ``collinear divergent'' (the
quotation marks indicate that the divergence is actually regularized by
the photon virtuality) cross section and factorization is broken. In
order to avoid such pathologies, we never work in the laboratory frame and
we impose, to insure the applicability of perturbation theory, a minimum
value on the transverse momentum of particles measured in the CM$^*$
frame.

For our calculation we use the following cuts, in agreement with
the H1 detector constraints. The lepton inelasticity is limited by
0.1 $\le y \le$ 0.7 and the Bjorken variable by 1. $\le  10^4\
x_{Bj} \le$ 5. The  forward $\pi^0$ is observed in the laboratory
in the domain  5$^{\circ} \le \theta_\pi^{\rm lab} \le$ 25$^{\circ}$
with a scaled energy .01 $\le x_\pi = E_\pi^{\rm lab} / E_p^{\rm
lab} \le$ 1. Furthermore we impose, in the CM$^*$ frame,
${p^*_{\bot}}^{\pi} \ge$ 4 GeV and  ${p^*_{\bot}}^{\rm jet} \ge
3.5$ GeV. This last cut is defined to avoid too large higher-order
corrections in some regions of phase space. The range of $Q^2$ is
limited to 50 GeV$^2$ and, in our study, we vary the lower bound
from 2 GeV$^2$ upward.

Before presenting the results it is worthwhile to discuss the
constraints imposed by the above cuts. First, let us recall the
necessary conditions for the BFKL dynamics to be relevant: one
should maximize the rapidity extent of the photon-parton system.
If $\widehat W$ be the energy of this system, we should maximize
the quantity (we recall the relation $y S = Q^2 / x_{Bj}$):
\begin{equation}
 {\widehat{W}^2 \over \sqrt{Q^2} p^*_{\perp}} =
{\sqrt{Q^2} \over  p^*_{\perp}}\ {(x_1 - x_{Bj}) \over {x_{Bj}}}
\nonumber \\
 =  {\sqrt{y S} \over p^*_{\perp}} \ {(x_1 - x_{Bj})\over \sqrt{x_{Bj}}}
\label{eq:1} 
\end{equation} 
where $S$ is the total electron-proton
energy squared, $x_1$ the fraction of the proton momentum carried
by the struck parton and $p^*_{\perp}$ the transverse momentum of the
final state parton. As is well known, this is achieved by using large $y$ and
small $x_{Bj}$, for $p^*_{\perp}$ not too large. Considering the Born
term kinematics with two hard partons in the final state, we can
derive the following relation in the CM$^*$ frame
\begin{equation} 
\exp({2 \eta^*_\pi-\Delta \eta^*}) = x_1 - {Q^2
(1-x_1) \over yS-Q^2} = {x_1 - x_{Bj} \over 1 - x_{Bj} } \sim x_1
- x_{Bj}\ , 
\label{eq:2} 
\end{equation} 
where $\eta^*_\pi$ is the
rapidity of the observed forward pion in the CM$^*$ frame and $\Delta
\eta^*$ the rapidity gap between the pion and the jet. The angular
conditions on the forward pion constrain the pion rapidity in the
laboratory,  $1.50 \leq \eta_{\pi}^{\rm lab} \leq 3.13$, or
approximately{\footnote{In the CM$^*$ frame, the pion rapidity is
approximately ($Q^2 \ll p_{\bot}^{*2}$) given by $\eta_{\pi}^* =
\eta_{\pi}^{\rm lab} - 1.7 + {1 \over 2} \ln y$ and we have used
${1 \over 2} \ln y \sim -.46$ for $y \sim .4$}} in the CM$^*$ frame
$-.65 \leq \eta_{\pi}^* \le 1$. When looking at the pion-jet
correlations, if we further impose the condition of a forward going
jet, $\Delta \eta^*$ decreases and $x_1$ increases. Therefore the
condition $\Delta \eta^*$ small corresponds to a forward jet implying
a large value of $x_1$ where the proton structure function is
suppressed. Another relation is (for $x_{Bj} \ll x_1$)
\begin{equation} 
p^2_\perp \cosh^2{\Delta\eta^* \over 2} = {Q^2 \over 4} {(x_1 - x_{Bj}) 
\over  x_{Bj}} = {x_1 y S - Q^2 \over 4}  \sim x_1 y \ {S \over 4} \ .  
\label{eq:3} 
\end{equation}


\noindent showing that at large values of $\Delta\eta^*$ correspond
large values of $y$ (and small values of $x_1$ by eq.~\ref{eq:2}) which are
cut by the condition $y < .7$.

\begin{figure} 
\begin{center}
\includegraphics[width=5.5in]{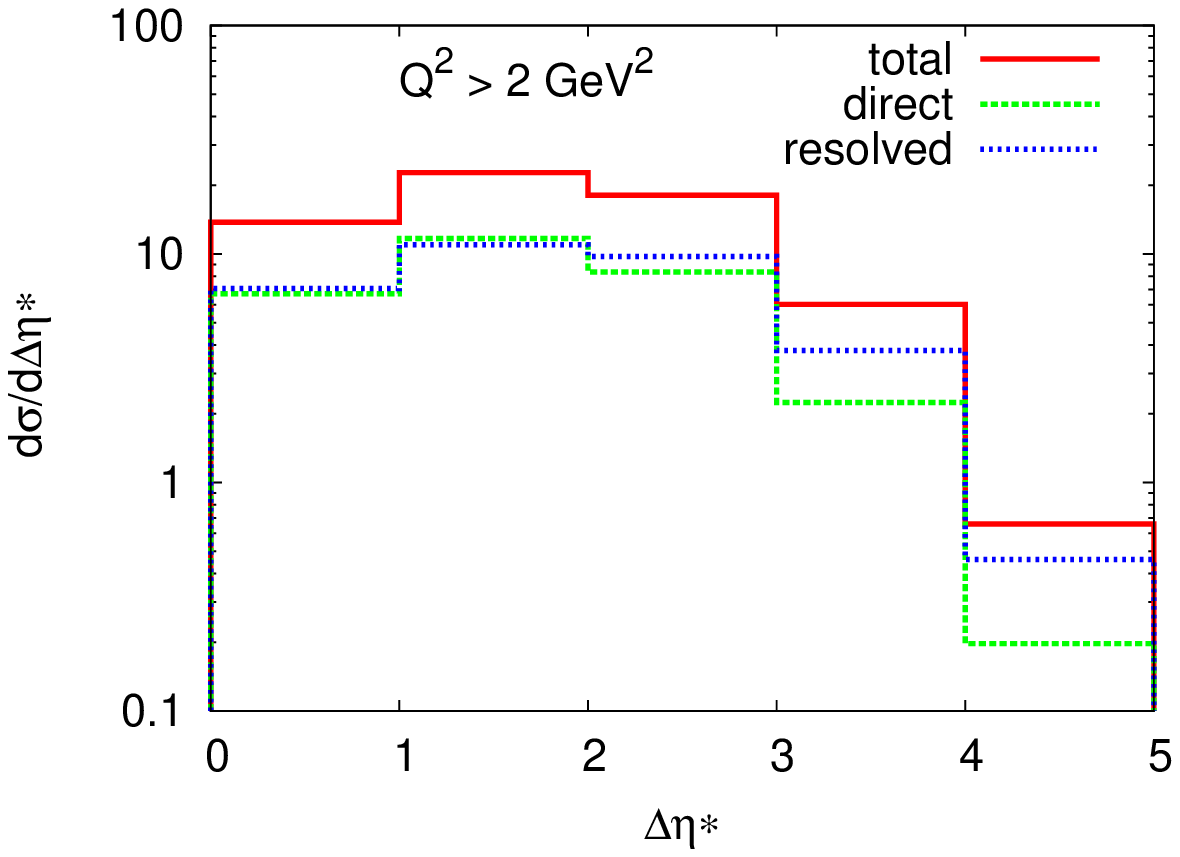}
\includegraphics[width=5.5in]{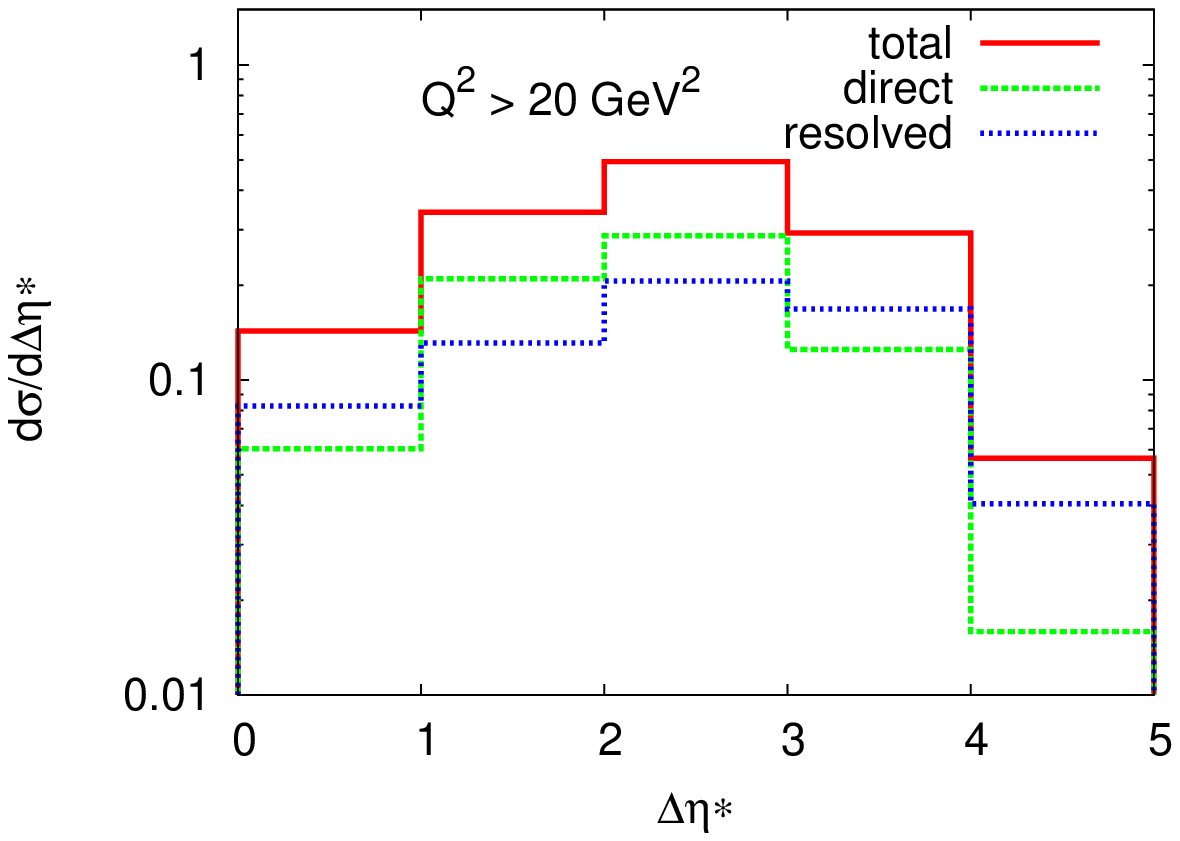} 
\end{center}
\caption {Variation of the cross section as a function of the
rapidity gap between the forward pion and the largest $p_\perp$
jet.} \label{fig:4} 
\end{figure}

These features can be clearly seen in Figs.~\ref{fig:4} where the jets are
defined as those with the largest transverse momentum. No rapidity plateau
appears in the $\Delta\eta^*$ dependence of the cross section and the
maximum of the cross section moves to larger values of $\Delta \eta^*$
when $Q_{min}^2$, and correspondingly  $y$, increases, as expected. In these
figures it is worth noting that the resolved cross section, the relative
weight of which decreases with increasing $Q^2$, is less affected by the
kinematics constraints discussed above in the Born case. Because it
involves an extra convolution with the parton distribution functions in the
photon, it is flatter than the direct contribution.

Some of the kinematical issues can also be understood by writing $y$ and $x_1$ 
in terms of the rapidity of the final partons $y_1$ and $y_2$. In a frame
in which the proton and photon are collinear ($P$ is the proton momentum in
this frame)

\begin{equation}
y = \frac{p^*_\perp}{\sqrt{S}}\frac{2P}{\sqrt{S}}(e^{-y_1}+e^{-y_2})
\label{y}
\end{equation}
and
\begin{equation}
x_1 = \frac{p^*_\perp}{\sqrt{S}}\frac{\sqrt{S}}{2P}(e^{y_1}+e^{y_2}) +
x_{Bj}
\label{x1}
\end{equation}
These equations directly give the variations of $x_1$ and $y$ in terms of the
rapidity of the final state jets and thereby the dependence of the parton
distributions of the proton and the photon on these rapidities. Thus, a second
forward parton corresponds to a larger value of $x_1$ whereas a backward 
parton corresponds to a larger value of $y$. These kinematical domains 
are thus suppressed by the large $x_1$ and large $y$ behaviour of the proton
and photon distribution functions respectively. Moreover, in the resolved case,
eq.(\ref{y}) is modified by multiplying the $y$ on the LHS by an extra
convolution $z$ due to the further splitting of the photon into a 
$q{\bar q}$ pair. Thus, for the direct case, 
$y_2$ is determined by $Q^2$, $y$
and $y_1$ which also determines the cross section. In the resolved case,
because of the extra convolution as explained above, $y_2$ is shifted to the 
right, and hence the whole resolved cross section, after averaging over
$Q^2$, $y$, $y_1$ and $z$ is also shifted to the right.   


There is a marked difference between the rapidity dependence of
these cross sections and that of the dijet measured in
ref.~\cite{Aktas:2004px}. There, the dijet cross section is shown
as a function of the rapidity gap between the two largest $p_{\bot}$
jets measured in the CM$^*$ frame but no special requirement is
imposed on the rapidity of one of the jets. The cross section is
maximum at $\Delta \eta^* = 0$ and slowly decreases with increasing
rapidity gap. The difference in shape with our results is due to
the requirement that the $\pi$ be observed in the forward direction,
in the laboratory frame, which introduces strong kinematical
constraints which distort, in our case, the rapidity dependence of
the cross section. This is confirmed by later studies of the H1
Collaboration~\cite{Aktas:2005up} where the rapidity dependence
of the cross section between a forward jet and a di-jet system of
small rapidity separation, is shown. A peak is observed at $\Delta
\eta^* \sim 1$ with the cross section rapidly falling on either
side. Let us note that this observable does not allow us to cover
a large rapidity gap domain (about 2 units for the jet-dijet
correlation instead of 5 in the particle-jet case).

\begin{figure} \vspace{9pt} 
\begin{center}
\includegraphics[width=5.5in]{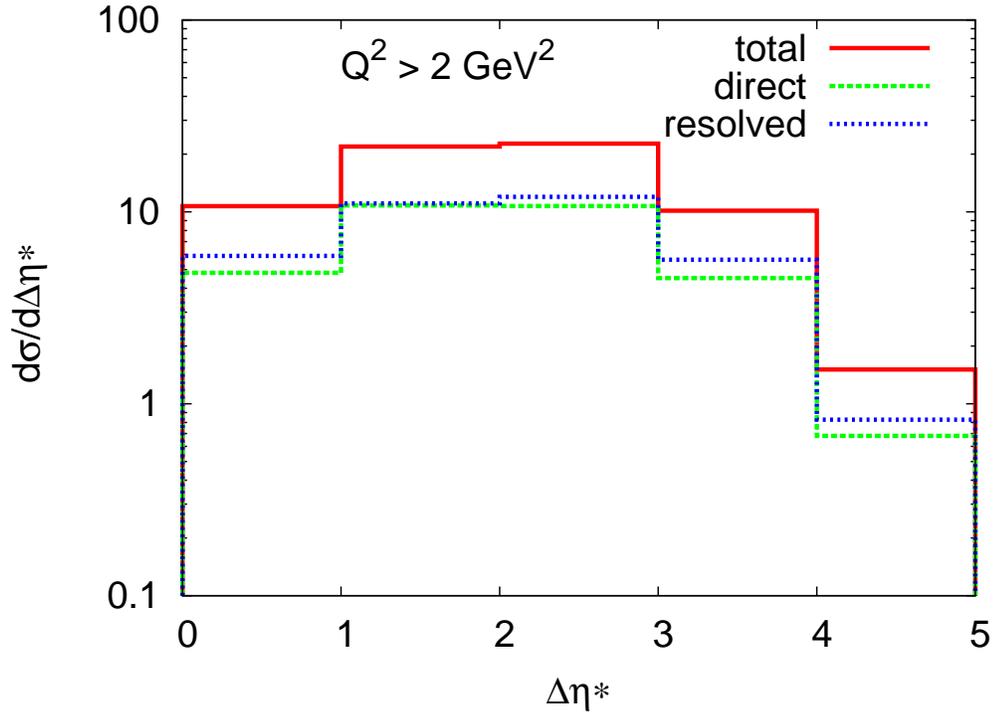}
\includegraphics[width=5.5in]{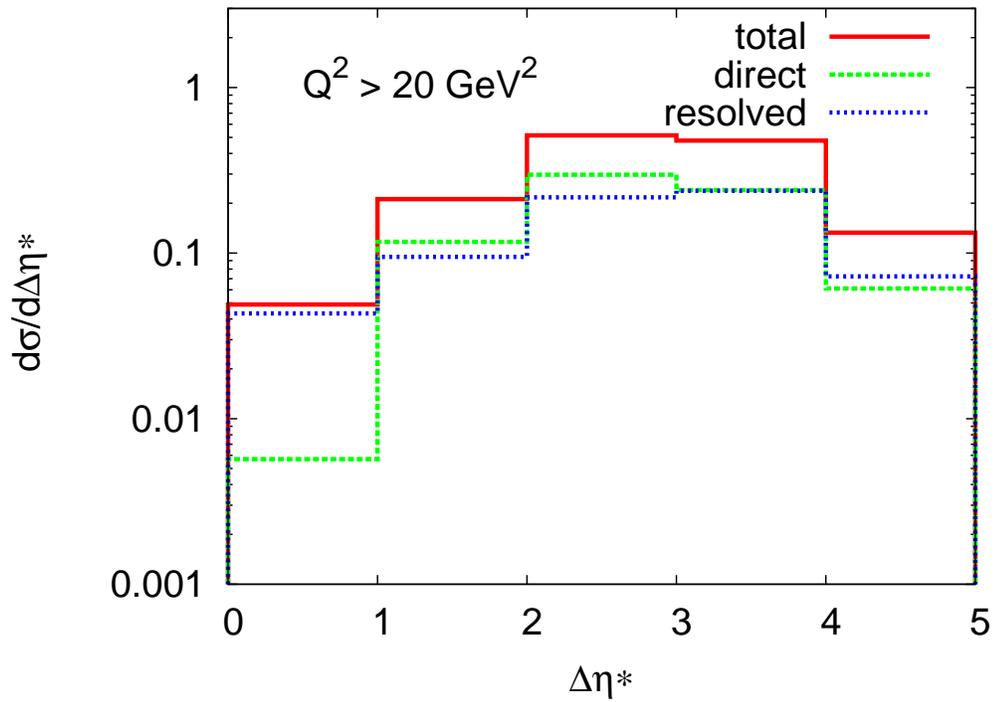} 
\end{center} 
\caption
{Variation of the cross section as a function of the rapidity gap
between the forward pion and the jet with the largest rapidity gap
with the pion.} 
\label{fig:5} 
\end{figure}

We now present in Figs.~\ref{fig:5} the cross section $d\sigma
/d\Delta \eta^*$ where $\Delta \eta^*$ is the rapidity gap, still
measured in the CM$^*$ frame, between the forward pion and the jet
(with a minimum transverse momentum as specified above) which has
the largest rapidity gap with the pion. The same general features
as in Figs.~\ref{fig:4} are observed with however a somewhat flatter
behavior than before at medium rapidity gaps and a slower decrease
at large gaps. This is related to the fact that the most backward
jet is not necessarily the largest $p_\perp$ one. This observable
can be compared to the one discussed by the D\O\ collaboration where
two jets with the largest separation are considered.

Let us now discuss some features of the NLO calculation.  The HO
corrections to the Born cross sections can be important in some
kinematical domains. Consider, as an example, the ``largest-rapidity-gap''
case and the value $Q_{min}^2 = 10$~Gev$^2$. For the direct
contribution the ratio HO/Born is equal to 1/3 in the range
$\Delta\eta^* \approx$ 0-1 while, in the range $\Delta \eta^* =$
4-5, it reaches 20. This large variation is due to the opening of
a new channel, namely $\gamma^*g \to q\bar{q} g$ (cf. Fig.~\ref{fig:3}(a)),
which becomes important at large values of $\Delta \eta^*$. The
variation of the ratio HO/Born is weaker in the resolved case: it
varies from 1/3 to 4 for $\Delta \eta^*$ increasing from the range
0-1 to the range 4-5. This increase is due to the larger phase space
available for the third jet when $\Delta\eta^*$ increases. This
behavior of the ratio  HO/Born partly explains those of the average
$<\cos (\pi - \phi )>$ which we now consider.

\begin{figure} 
\vspace{9pt} 
\begin{center}
\includegraphics[width=4.5in]{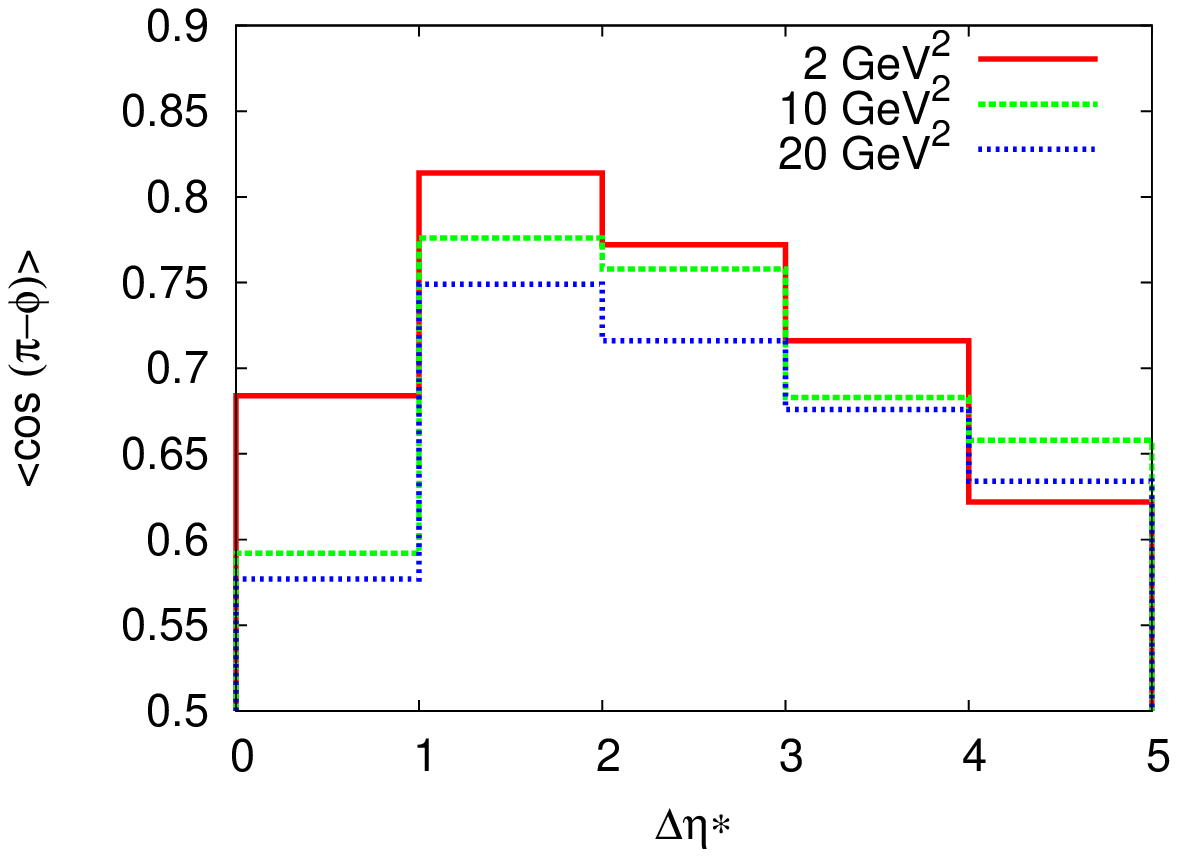} 
\end{center} 
\caption
{Average $<\cos (\pi - \phi )>$ as a function of the rapidity gap
between the forward pion and the most backward jet. The different
curves correspond to the minimum values of $Q^2$ used in the
calculation.} 
\label{fig:6} 
\end{figure}

\begin{figure} 
\vspace{9pt} 
\begin{center}
\includegraphics[width=4.5in]{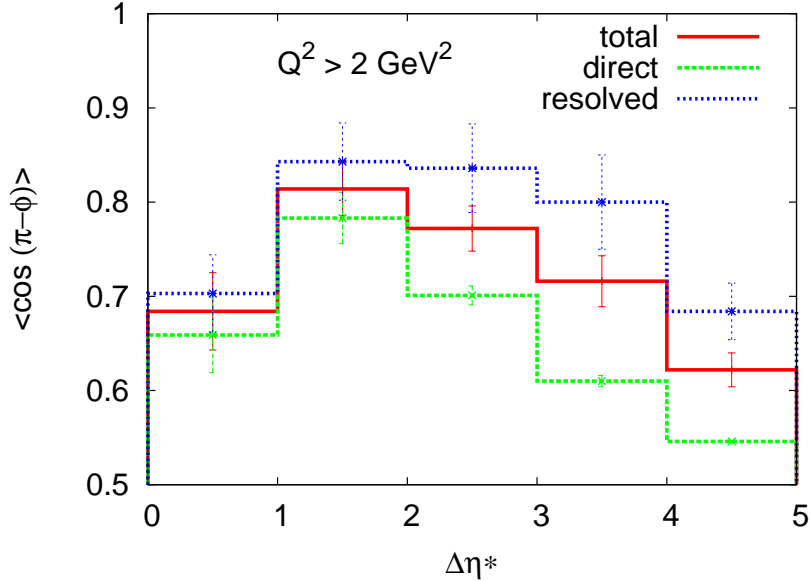} \end{center} \caption {Detail
of the components of $<\cos (\pi - \phi )>$ as a function of the
rapidity gap between the forward pion and the most backward jet
for $Q^2> 2$ GeV$^2$.} \label{fig:7} 
\end{figure}

Figs.~\ref{fig:6},~\ref{fig:7}\footnote{Note that the errors in Fig. 7 are 
large primarily because of the errors
in the resolved case. As usual the errors in the cross section and the
$\cos \phi$ weighted cross section have been added in quadratures to obtain
the errors in the $<\cos (\pi - \phi)>$ calculation.} give the dependence 
of the average $<\cos (\pi -
\phi )>$ as a function of the rapidity gap under the same conditions as those of
Fig.~\ref{fig:5}. The decrease of $< \cos (\pi - \phi )>$ when $\Delta \eta^*$
increases is explained by the increase of the phase space available for the
final jets in the $2\to 3$ subprocesses.  Going into the details of the
underlying mechanisms we see in Fig.~\ref{fig:7} that the direct component shows
a more rapid decorrelation as $\Delta\eta^*$ increases because of the rapid
growth of the HO contribution to this process. Comparing with the D\O \ data of
Fig.~\ref{fig:1}, we note that the overall size of $<\cos (\pi - \phi )>$ is
smaller for the same value of $\Delta\eta^*$ but that the slope is rather
similar at large $\Delta \eta^*$. 

Coming back to Fig.~\ref{fig:6} it is interesting to note the
decrease of $< \cos (\pi - \phi )>$ when $Q_{min}^2$ increases.
This corresponds to the $2 \to 3$ subprocesses becoming more isotropic.  
Finally the dip at $\Delta \eta^*$ close to zero can be explained in the
following way.  First we note that in this $\Delta \eta^*$ range, the HO
corrections are smaller than the Born term. In the direct case, they even become
negative when $Q_{min}^2$ increases. In fact, this means that there are large
compensations between the negative quasi $2\to 2$ HO contributions (virtual and
collinear) and the positive contributions coming from the $2\to 3$ processes.
When the $2\to 3$ contributions are weighted by $\cos (\pi - \phi )$, the
(weighted) HO contributions strongly decreases, which causes a decrease of $<
\cos (\pi - \phi )>$. For higher values of $\Delta \eta^*$, this effect is less
pronounced. Let us also note that the resolved component is dominant in 
this bin for $Q^2\ >\ 10\ GeV^2$. Without HO corrections to this component, 
we would have obtained $<\cos(\pi-\phi)>\sim 1.0$ 

The behavior of $< \cos (\pi - \phi )>$ with $\Delta \eta^*$ is
strongly dependent on kinematical constraints, which are exactly respected in
our NLO calculation, and this effect is specially strong near $\Delta \eta^*\sim 0$.
On the other hand the BFKL approach does not usually implement total energy
momentum conservation\footnote{For a discussion see 
ref. \protect{\cite{Orr:1997im}}.}
and we expect quite different predictions.

\mysection{Conclusion} 
\hspace*{\parindent} 
In this work we compare the azimuthal decorrelation effects, as a function of
the rapidity gap between two jets at the Tevatron and between a forward
$\pi^0$ and a jet at HERA, using a next to leading order
formalism. In the case of hadronic collisions, the calculated decorrelation
effect is slightly slower than the data but it is still compatible with them
taking into account statistical and systematic errors. In the deep inelastic
case, the decorrelation is somewhat stronger than in the $p\ \bar p$ case,
increasing with $Q^2$, and we note interesting kinematical effects, related to
the requirement of a forward particle trigger. In particular, it introduces a
dip in the cross section at small rapidity gap, between the hadron and the jet.
Dynamical effects and kinematical constraints also lead to very specific 
variations of $<\cos(\pi-\phi)>$ with $\Delta\eta^*$.

\end{document}